\documentclass[aps,prd,twocolumn,groupedaddress,showpacs]{revtex4}
\usepackage{graphicx}

\begin{document}

\title
{
\begin{flushright}
{\normalsize  CU-TP-1138}\\
\end{flushright}
Quark-gluon vertex with an off-shell \(O(a)\)-improved chiral
fermion action}
\author{Huey-Wen Lin}
\email{hwlin@theory1.phys.columbia.edu} \affiliation{Physics
Department, Columbia University, New York, NY 10027, USA}


\begin{abstract}
We perform a study the quark-gluon vertex function with a quenched
Wilson gauge action and a variety of fermion actions. These
include the domain wall fermion action (with exponentially
accurate chiral symmetry) and the Wilson clover action both with
the non-perturbatively improved clover coefficient as well as with
a number of different values for this coefficient. We find that
the domain wall vertex function behaves very well in the large
momentum transfer region. The \emph{off-shell} vertex function for
the \emph{on-shell} improved clover class of actions does not
behave as well as the domain wall case and, surprisingly, shows
only a weak dependence on the clover coefficient $c_{\rm SW}$ for
all components of its Dirac decomposition and across all momenta.
Including off-shell improvement rotations for the clover fields
can make this action yield results consistent with those from the
domain wall approach, as well as helping to determine the
off-shell improved coefficient $c_q^\prime$.
\end{abstract}


\maketitle \vspace{ -0.3in}
\section{Introduction}

The study of gauge-fixed quark and gluon correlation functions on
the lattice is a useful tool to understand the non-perturbative
behavior of QCD. Within lattice QCD one may hope that the
quark-gluon vertex in particular may be a momentum dependent probe
that displays discretization effects beyond those seen in the free
field case. Indeed one may hope that knowledge of the quark-gluon
vertex allows the determination of improvement coefficients
without resort to Ward identities. This may allow the
non-perturbative \(O(a)\) improvement obtained in the light quark
limit for clover fermions\cite{NP_CLOVER} using the PCAC relation
to be extended to the Fermilab action\cite{Fermilab_action} at
arbitrary quark mass.

Beyond lattice QCD knowledge of the non-perturbative $n$-point
functions gives the only theoretical access to the deeply
infra-red sector of the theory and may be used to provide
non-perturbative information for model calculations. Further we
can obtain the non-perturbative running of a physically defined
coupling, and may be able to use these non-perturbative
correlation functions in the Dyson-Schwinger equations. DSEs are
of qualitative and quantitative importance in understanding
dynamical chiral symmetry breaking and confinement, and
simulations of lattice QCD also provide access to QCD's Schwinger
functions. Previous studies of the quenched theory have yielded
gluon\cite{quenched_g_prop} and clover quark\cite{quark_prop}
propagators and the quark-gluon vertex on the lattice has also
been studied extensively\cite{qgv}. These previous studies of the
vertex were limited to the on-shell improved, non-chirally
symmetric clover action, and post-simulation field rotations were
used to implement off-shell improvement.

The domain wall chiral fermion action\cite{DWF_prehistory} is
automatically \(O(a)\)-improved, and domain wall fields should
greatly aid studies of off-shell correlation functions. The
off-shell quark propagator has been previously and comprehensively
studied for the domain wall fermion action\cite{RBC_NPR}, and
renormalization constants were determined for many fermion
bilinears in the chiral limit for the Wilson and DBW2 gauge
actions. The off-shell properties of the propagator were found to
be remarkably continuum-like for this action, and one expects that
this would also be seen for the domain wall vertex function.

In this paper we also use this more advanced domain wall fermion
action, which is chirally symmetric (bar exponentially small
terms), and demonstrate that such improvement is indeed seen in
our calculations of the quark-quark-gluon three-point function and
the vertex function.

Further, one might expect that a Dirac structure decomposition of
the vertex function could be used to match the clover action (with
field rotations) to the domain wall action. This could provide a
mechanism for fixing the clover coefficient non-perturbatively
without the imposition of the PCAC relation, and it was hoped that
this would allow a non-perturbative clover coefficient
determination for all quark masses\cite{hwlin04}. However we find
weak dependence on this coefficient after absorbing a trivial mass
renormalization and instead find the correlation function is most
sensitive to the $c_q^\prime$ improvement parameter.

The structure of this paper is as follows:
Section~\ref{SecDefinitions} defines the notation and lattice
correlation functions measured to obtain gauge-fixed two-point and
three-point functions. Section~\ref{SecParams} lists details of
the simulation parameters. Section~\ref{SecGaugeFix} presents the
gauge-fixing strategy used.  Sections \ref{SecFermionProp} and
\ref{SecGluonProp} present simulation results of the quark and
gluon two-point functions. Section ~\ref{SecVertex} presents
simulation results of the quark-quark-gluon three-point function
and the truncated vertex. Section ~\ref{SecDiscuss} discusses the
off-shell improvement and contrasts the DWF and clover cases.
Finally, Section~\ref{SecFuture} presents a summary, conclusions,
and future outlook for this program.

\section {Non-perturbative Green functions: Definition}
\label{SecDefinitions}
\subsection {Off-shell \(O(a)\) improvement}
\label{SecOffShellImpro} The on-shell \(O(a)\)-improved Wilson
clover action\cite{clover} can be written as
\begin{eqnarray}
S  &=&  a^4\sum_{x, x^\prime} \bar{\psi(x)}(D_{c}+m)_{x x'}\psi(x^\prime)\\
(D_{c} + m )_{x x^\prime} &=&  m -\frac{1}{2a}\sum_{\mu} \left\{
(1-\gamma_{\mu})U_{x, \mu} \delta_{x+\hat\mu,x^\prime}
\vphantom{U^\dagger_{x^\prime, \mu}}\right.
\nonumber\\
&+& \left. (1+\gamma_\mu) U^\dagger_{x^\prime, \mu}
\delta_{x-\hat\mu,x^\prime} \right\}
\nonumber\\
&+& \left\{\frac{4}{a}+%
\frac{ic_{\mathrm{SW}}}{4} \sum_{\mu,\nu}\sigma_{\mu\nu} F_{\mu
\nu, x x^\prime}
\right\}\delta_{x,x^\prime} \\
F_{\mu \nu , x x^\prime}& = &\frac{1}{2a^2} \left\{U_{x,
\mu}U_{x+\hat\mu, \nu}U^\dagger_{x+\hat\nu, \mu}U^\dagger_{\nu}
\right. \nonumber\\
& -&  U^\dagger_{x-\hat\nu, \nu}U^\dagger_{x-\hat\nu-\hat\mu,
\mu}U_{x-\hat\nu-\hat\mu, \nu}U_{x, \mu}\nonumber\\
&+& U_{x, \nu}U^\dagger_{x+\hat\nu-\hat\mu,
\mu}U^\dagger_{x-\hat\mu, \nu}U_{x-\hat\mu, \mu}\nonumber\\
&-& \left. U_{x, \mu}U^\dagger_{x-\hat\nu+\hat\mu,
\nu}U^\dagger_{x-\hat\nu, \mu}U_{x-\hat\nu, \nu}
\right\}.
\end{eqnarray}

To \emph{off-shell} improve the physical quantities calculated
from this Wilson clover action, we keep the action unchanged but
perform post-simulation rotations\cite{Martinelli_offShell} to
improved quark fields of the form
\begin{eqnarray}
\psi \rightarrow \hat{\psi} &=& Z_q^{\frac {1}{2}} \left[1+a
c_q^\prime (D_{c}+m) + a c_\mathrm{NGI} \gamma_{\mu}
\partial_{\mu}
\right]\psi\nonumber\\
\overline{\psi} \rightarrow \hat{\overline{\psi}} &=& 
Z_q^{\frac {1}{2}} \overline{\psi} \left[1+a{c_q}^\prime
(-\!\!\stackrel{\leftarrow}{D_c} + m) - ac_\mathrm{NGI}
\gamma_{\mu}\stackrel{\leftarrow}{\partial_{\mu}}\right].
\label{eq:off_shell_improv_psi}
\end{eqnarray}

The improved quark propagator $S_I(x) \equiv \langle \hat{\psi}(x)
\hat{\overline{\psi}}(0) \rangle $  can then be expressed in terms
of the unimproved quark propagator. In momentum space, we have:
\begin{eqnarray}
{S_I}(p) &=& {Z_q}
\left[\left(S_L(p)+2a c_q^\prime\right) \vphantom{\frac{pa}{a}} \right. \nonumber \\
&+& \left. a c_\mathrm{NGI}\left\{
i\gamma_{\mu}\frac{pa}{a},S_L(p) \right\} \right] + O(a^2)
\end{eqnarray}
where $S_L$ is the unimproved quark propagator.  The constant
$c_\mathrm{NGI}$ has been previously demonstrated to be zero at
tree level\cite{Martinelli_offShell} and small for light quark
masses\cite{C_ngi}.

The domain wall fermion (DWF) action\cite{DWF_prehistory} is
\begin{equation}
  D_{x,s; x^\prime, s^\prime} = \delta_{s,s^\prime}
    D^\parallel_{x,x^\prime} + \delta_{x,x^\prime} D^\bot_{s,s^\prime}
\label{eq:d}
\end{equation}
\begin{eqnarray}
D^\parallel_{x,x^\prime} \hspace{-2.5mm}&=&\hspace{-2mm}
  {1\over 2} \sum_{\mu=1}^4 \left[ (1-\gamma_\mu)
  U_{x,\mu} \delta_{x+\hat\mu,x^\prime} + (1+\gamma_\mu)
  U^\dagger_{x^\prime, \mu} \delta_{x-\hat\mu,x^\prime} \right]
  \nonumber \\
  \hspace{-2mm}& +& \hspace{-2mm} (M_5 - 4)\delta_{x,x^\prime}
\label{eq:D_parallel}
\end{eqnarray}
\begin{eqnarray}
D^\bot_{s,s^\prime}
    \hspace{-1mm} &=& \hspace{-1mm} {1\over 2}\Big[(1-\gamma_5)\delta_{s+1,s^\prime}
                 + (1+\gamma_5)\delta_{s-1,s^\prime}
                 - 2\delta_{s,s^\prime}\Big] \nonumber\\
   \hspace{-1mm}  &-& \hspace{-1mm}{m_f\over 2}\Big[(1-\gamma_5) \delta_{s, L_s-1}
       \delta_{0, s^\prime}
      +  (1+\gamma_5)\delta_{s,0}\delta_{L_s-1,s^\prime}\Big],
      \nonumber\\
\label{D_perp}
\end{eqnarray}
where $s$ and $s^\prime$ lie in the range $0 \le s,s^\prime \le
L_s-1$, $M_5$ is the five-dimensional mass, and $m_f$ directly
couples the two domain walls at $s=0$ and $s = L_s - 1$. The DWF
action is \(O(a)\) off-shell improved due to its exponentially
accurate chiral symmetry\cite{RBC_NPR}, and no further improvement
in the action or quark fields is performed.
\subsection {The gluon field}
The lattice gluon field $A_{\mu}^a$ is defined by
\begin{eqnarray}
A_{\mu}^a(x+\hat{\mu}/2)\tau^a&=
&\frac{1}{2ig_0}\left[\left(U_{\mu}(x)-
U_{\mu}^\dag(x)\right) \vphantom{\frac{1}{3}} \right. \label{eq:gluon_field} \nonumber\\
\!\! \!&-&\!\!\! \left. \frac{1}{3} \mathrm{Tr} \left(U_{\mu}(x)-
U_{\mu}^\dag(x)\right)\right] + O(a^3),
\end{eqnarray}
where \(a\) is the color index, $\mu$ is the Lorentz index, $g_0$
is the bare coupling constant, and $\mathrm{Tr}\{ \tau^a \tau^b
\}= \frac{1}{2}\delta_{ab}$.

The gluon propagator (in Landau gauge) is
\begin{eqnarray}
D_{\mu\nu}^{ab}(q) = \frac{1}{V}\langle A_{\mu}^a(q) A_{\nu}^b(-q)
\rangle \simeq \delta_{ab} P_{\mu\nu} D(q^2),
\end{eqnarray}
where $D(q^2) = \frac{1}{24} \sum_{\mu,a} D_{\mu\mu}^{aa}(q) $ and
$P_{\mu\nu} = \delta_{\mu\nu} - \frac{q_{\mu}q_{\nu}}{q^2}.$

\subsection {The quark-gluon vertex function}
The quark-gluon vertex function can be calculated from the
following three-point function\cite{qgv}:
\begin{eqnarray}
&&\hskip -0.2in V_{\mu}^a(x,y,z)_{\alpha \beta}^{ij} = \langle
S_{\alpha\beta}^{ij}(x,z) A_{\mu}^a(y)\rangle.
\end{eqnarray}
where the lattice gluon field $A_{\mu}^a$ is defined in
Eq.~\ref{eq:gluon_field}. The Fourier-transformed, amputated and
transverse-projected vertex function in momentum space for Landau
gauge is given by
\begin{eqnarray}
&&\hskip -0.2in \Lambda_{\mu}^{P{} , a,{\rm lat}}(p,q)_{\alpha
\beta}^{ij} = \Lambda_\mu(p,q)_{\alpha \beta} \tau^a_{ij} =
P_{\mu\nu} \Lambda_{\nu}^{
a,{\rm lat}}(p,q)_{\alpha \beta}^{ij}\nonumber\\
&&\hskip -0.05in = \langle {S}(p)\rangle^{-1} P_{\mu\nu}
V_{\nu}^a(p,q)_{\alpha \beta}^{ij}\langle {S}(p+q)\rangle^{-1}
\langle D(q)_{}\rangle^{-1}
\end{eqnarray}
where the projected vertex function is introduced since the gluon
propagator is singular for general momenta, posing problems for a
full amputation. Note that the $S$ means $S_L(p)$ for the bare
vertex and $S_I(p)$ for the off-shell improved clover quark field.

In the case of $q=0$ the gluon propagator is no longer
proportional to the projection $P_{\mu\nu}$, but rather is
\begin{eqnarray}
&&\hskip -0.2in \Lambda_{\mu}^{ a,{\rm lat}}(p,q = 0)_{\alpha
\beta}^{ij} \nonumber\\
&&\hskip -0.1in = \langle {S}(p)\rangle^{-1} P_{\mu\nu}
V_{\nu}^a(p,q=0)_{\alpha \beta}^{ij}\langle {S}(p)\rangle^{-1}
\langle D(0)_{}\rangle^{-1}.
\end{eqnarray}
The non-perturbative gluon propagator in a finite volume remains
well defined and the same amputation process is used.

\section{Simulation parameters}
\label{SecParams} All of the calculations in this paper were
performed with the Wilson gauge action\cite{wilson_g} in the
quenched approximation at $\beta = 6.0$ , $a^{-1} = 1.922(40)$ GeV
with a $16^3 \times 32$ lattice size\cite{dwf_00}. The simulation
was performed using a heatbath algorithm with $2000$
thermalization sweeps and $500$ measurements, separated by $2000$
sweeps each.

The fermion actions used are DWF and the Wilson clover action with
various parameters. The details are listed in
Table~\ref{tab:params}, and we note that the pseudoscalar masses
for three of the clover action combinations were (somewhat
approximately) matched to the domain wall pseudoscalar mass using
a single, low-statistics pass.

\begin{table}[hbt]
\caption{\label{tab:params}Simulation parameters }
\begin{tabular}{c|c|c}
 Label & Parameters & PS meson \\
\hline %
  DWF & $M_5 = 1.8$, $L_s = 16$, &  442(9) MeV\\
  &   $ma = 0.0125$ &    \\
\hline %
$c_\mathrm{SW}=1.769$ & $\kappa = 0.1346$, $c_\mathrm{SW} = 1.769$ & 470(10) MeV\\
$c_\mathrm{SW}=1.5$   & $\kappa = 0.1382$, $c_\mathrm{SW} = 1.5$   & 462(11)  MeV\\
$c_\mathrm{SW}=2.0$   & $\kappa = 01313$,  $c_\mathrm{SW} = 2.0$   & 452(13) MeV\\
$c_\mathrm{SW}=1.479$ & $\kappa = 0.1370$, $c_\mathrm{SW} = 1.479$ & 796(8) MeV\\
 \hline %
\end{tabular}
\end{table}

We use periodic boundary conditions for both the spatial and
temporal directions to calculate both the fermionic and gluonic
propagators. The momenta used for Fourier transformation are
$p_\mu = 2\pi n_\mu /L_\mu$, with $|n_{\{x,y,z\}}| \in \{0, 1,
2\}$ and $|n_{t}| \in \{1, 2\}$.

\section {Gauge fixing strategy}
\label{SecGaugeFix} Since we are interested in gauge-dependent
off-shell quantities we choose to fix to a convenient gauge,
namely Landau gauge, by maximizing the following functional:
\begin{eqnarray}
F[U_{\mu}(n), g(n)] = 1- \sum_{\mu, n}\mathrm{Re}\mathrm{Tr}
U_{\mu}^g(n),
\end{eqnarray}
where $U_{\mu}^g(n) = g(n)U_{\mu}(n)g(n+\hat{\mu})^\dag$ with
gauge transformation matrix $g(n)$. The gauge-fixing stopping
condition is
\begin{eqnarray}
\frac{1}{N_cV}\sum_{n}\mathrm{Tr}\left(B(n)B(n)^{\dag}\right) \le
10^{-8}
\end{eqnarray}
where $N$ is the number of total sites on the lattice and
\begin{eqnarray}
 B(n) &=& C(n)-C(n)^{\dag} - \frac{1}{3}\mathrm{Tr}\left(C(n)-C(n)^{\dag}\right), \\
 C(n) &=& \left(U_{\mu}^g(n)^\dag + {U^g_{\mu}(n -\hat{\mu}}^\dag)\right).
\end{eqnarray}

Landau gauge suffers from Gribov ambiguities\cite{gribov}, and due
care must be taken to ensure our results are not significantly
affected. We checked for the influence of Gribov copies on each of
our observables by gauge fixing in two ways. The first is
na\"ively fixing to Landau gauge in the usual way, and the second
involves fixing to Landau gauge from a deterministic starting
gauge orbit, by first fixing to maximal axial gauge\cite{Yuri}.
This is a common approach which gives a deterministic Gribov copy,
independent of the gauge of the initial lattice configuration.

\begin{figure}[hbt]
\vspace{0.8cm}
\includegraphics[width=0.9\columnwidth]{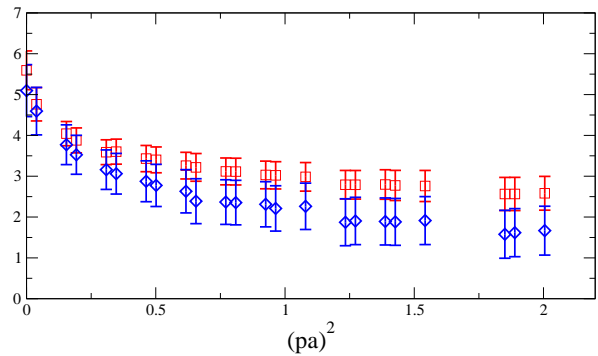}
\caption{ \label{asym_ver_MAG_comp} Comparison between Landau
gauge fixing with (squares) and without (diamonds) first fixing to
maximal axial gauge for the vertex function.}
\end{figure}

We found that the two different procedures agree for the gluon
propagator, have slight effects on quark propagators (decreasing
as the momentum increases) but give very different results in the
vertex calculation\cite{hwlin04}, as shown in
Figure~\ref{asym_ver_MAG_comp}. We believe the differences are
caused by Gribov copies and from now on, we will follow the
deterministic procedure for better control over this effect.

\section {Fermionic two-point Green function}
\label{SecFermionProp} We start from the two-point function used
for non-perturbative renormalization in the RI-MOM
scheme\cite{RI_NPR}. The quark and gluon propagators from quenched
lattice configurations have been well studied previously for both
fermion actions\cite{quark_prop,RBC_NPR}. However, these remain a
prerequisite for our exploratory study of the vertex function, and
our results must be defined and introduced prior to moving on to
the vertex function.

The bare quark field renormalization factor, $Z_q(p^2)$, is
calculated from
\begin{eqnarray}
 Z_q(pa) = \frac{-i}{12 \sum_{\mu} (p_{\mu}a)^2} \mathrm{Tr}\left(\sum_{\mu}
p_{\mu}a \gamma_{\mu} S_L^{-1} \right),
\end{eqnarray}
where, as we shall see, constraints on the above quantity in the
matching window allow $Z_q$  to be determined from a fit.

\begin{figure}[hbt]
\includegraphics[width=0.9\columnwidth]{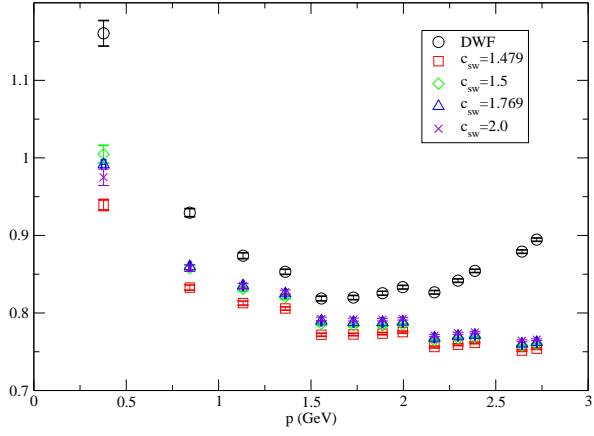}
\caption{ \label{Z2_comp_500} The quark field renormalization
factor $Z_q$ for different actions.  }
\end{figure}

In Figure~\ref{Z2_comp_500} we show the bare quark field
renormalization $Z_q((pa)^2)$ for each of the actions listed in
Table~\ref{tab:params} using the bare fields. We calculate the
$Z_q$ factor in the RI-MOM scheme by removing the expected
perturbative momentum dependence to define a scale invariant
quantity $Z_q^\mathrm{SI}$,
\begin{eqnarray}
Z_q^\mathrm{SI}((pa)^2) = Z_q((pa)^2)/ C_A((pa)^2),
\end{eqnarray}
and then fit the $(ap)^2$ dependence of this SI expression to the form
$f_1 + f_2 (pa)^2$ where the $f_2$ term absorbs lattice
artifacts.

\begin{figure}[hbt]
\includegraphics[width=0.9\columnwidth]{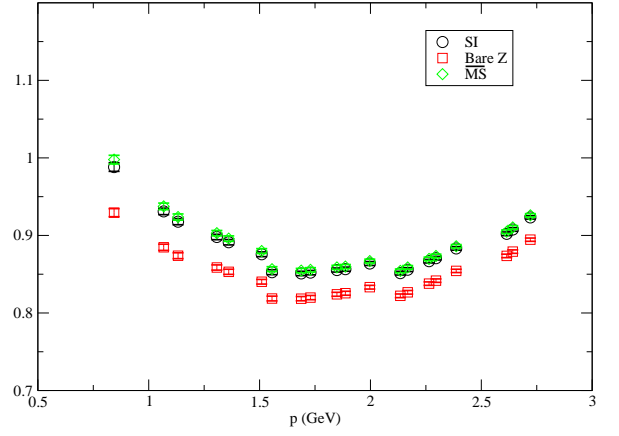}
 \caption{ \label{Z_running_DWF} The quark field
renormalization factor $Z_q$ for the bare, SI and
$\overline{\mathrm{MS}}$ schemes on our DWF data set. }
\end{figure}

From this we obtain the phenomenologically relevant
renormalization factor between the lattice and the
$\overline{\mathrm{MS}}$ scheme.
\begin{eqnarray}
\frac{Z^\mathrm{RI}}{Z^{\overline{\mathrm{MS}}}} = 1 +
\frac{\alpha_s}{4\pi}Z_0^{(1)\mathrm{RI}}+
\frac{\alpha_s^2}{(4\pi)^2}Z_0^{(2)\mathrm{RI}}
\end{eqnarray}
where definition and values for $C_A((pa)^2)$, $\alpha_s$,
$Z_0^{(1)\mathrm{RI}}$ and $Z_0^{(2)\mathrm{RI}}$ are well
known\cite{RBC_NPR}. Figure~\ref{Z_running_DWF} shows  $Z_q$ for
these three schemes for domain wall fermions.

Table~\ref{tab:Zq} shows our results for the various fermion
actions under consideration.  As one would expect, where previous
data exists for the same action, these results are consistent. In
particular, for the domain wall action we refer the reader to
Ref.~\cite{RBC_NPR} where the massless limit was taken.

\begin{table}[hbt]
\caption{\label{tab:Zq}Quark field renormalization factor for the
fitting range of $0.8 \le (pa)^2 \le 2.0$. }
\begin{tabular}{c|c|c}
 Label & $Z^\mathrm{SI}$ & $Z^{\overline{\mathrm{MS}}}$ \\
\hline %
DWF    &   0.802(9)  &  0.806(9) \\
$c_\mathrm{SW}=1.479$  &  0.825(6)  &  0.829(6) \\
$c_\mathrm{SW}=1.5$    &  0.840(6)  &  0.844(6)  \\
$c_\mathrm{SW}=1.769$  &  0.846(6)  &  0.850(6) \\
$c_\mathrm{SW}=2.0$    &  0.849(6)  &  0.854(6) \\
\hline %
\end{tabular}
\vspace{0.5cm}
\end{table}

\begin{figure}[!hbt]
\vspace{0.5cm}
\includegraphics[width=0.9\columnwidth]{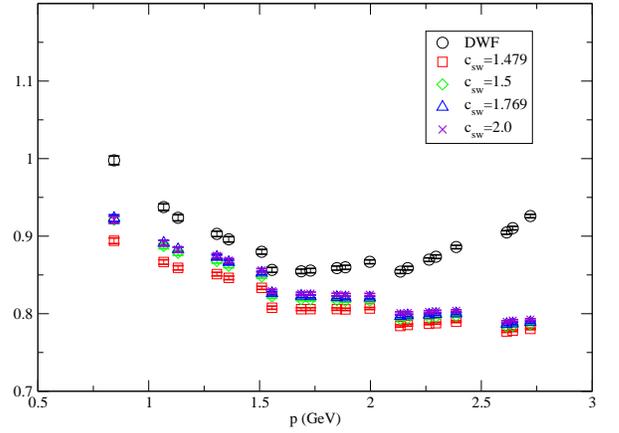}
 \caption{ \label{Z2_MS_comp_500} The quark field
renormalization factor $Z_q^{\overline{\mathrm{MS}}}((pa)^2)$ for
different actions.  }
\end{figure}

Following Ref.~\cite{Martinelli_offShell}, we expect that in the
large momentum region the scalar part of the quark propagator
behaves like
\begin{eqnarray}
\label{EqFitForm}
 \frac{1}{12}\mathrm{Tr}(S_L(p)) = C \times (pa)^2 + (-2c_q^\prime +
 2c_\mathrm{NGI}Z_q)\nonumber \\
 + \frac{Z_q Z_m
 m_I}{p^2} + O(\frac{1}{p^4}).
\label{EqFitForm}
\end{eqnarray}
As mentioned in Subsection~\ref{SecOffShellImpro},
$c_\mathrm{NGI}$ is zero at tree level and has been found to be
numerically very small in the light quark region\cite{C_ngi},
enabling us to take it as zero in our calculation. Therefore, we
can use this formula to determine $c_q^\prime$.

The resulting parameters, used later in this paper, are displayed
in Table~\ref{tab:fit_B}.  We note that the first two terms in
Eq.~\ref{EqFitForm} violate chrial symmetry and are not allowed in
the DWF case. Our values for the clover actions are plausible
given tree-level estimates of around 0.25.

\begin{table}[hbt]
\caption{\label{tab:fit_B} Fitting to the first three terms in Eq.
(\ref{EqFitForm}) with fitting range of $0.8 \le (pa)^2 \le 2.0$
produces the improvement coefficient $c_q^\prime$. }
 \begin{tabular}{c|c}
 \hline %
Label & $c_q^\prime$ \\
\hline %
 $c_\mathrm{SW}=1.479$ & $-0.247(19)$\\
 $c_\mathrm{SW}=1.5$   & $-0.258(19)$\\
 $c_\mathrm{SW}=1.769$ & $-0.257(19)$\\
 $c_\mathrm{SW}=2.0$   & $-0.252(19)$\\
            \hline %
\end{tabular}
\end{table}

\section {Gluonic two-point Green function}
\label{SecGluonProp} 
\begin{figure}[hbt]
\includegraphics[width=1.0\columnwidth]{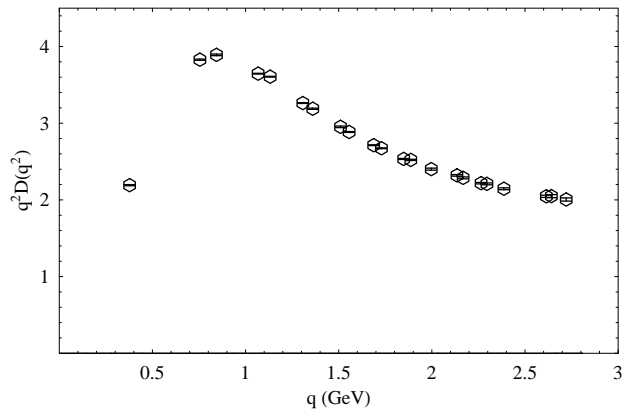}
\caption{ \label{Z3_500} The na\"ive RI-MOM scheme renormalization
constant $Z_3^{-1} = q^2 D(q^2)$ in Landau gauge.}
\end{figure}

There are many renormalization schemes that
may be applied to the gluon propagator, and one of the more common
in the literature is the MOM scheme where the gluon field
renormalization factor is constrained by requiring that, at some
fixed scale $\mu$, an effective tree level Coulombic form holds:
\begin{eqnarray}
Z_3^\mathrm{RI}(\mu) D(q^2;\mu)|_{q^{2}={\mu}^{2}} &=&
\frac{1}{\mu^2}.
\end{eqnarray}
The quantity $Z_3(q^2)^{-1} = q^2 D(q^2) $ is displayed in
Figure~\ref{Z3_500}, and we see that this condition leads to a
renormalization constant $Z_3$ that is highly sensitive to the
scale $\mu$ in the momentum regions currently accessible to
lattice QCD.

This scale dependence is an artifact of the renormalization
condition, and we pick a fixed renormalization point $a\mu=1$ and
use the one-loop form\cite{gluon_fit} of the running to present
the data in a manner that should be perturbatively
scale-invariant:
\begin{eqnarray}
\left(Z_3^\mathrm{SI}(q^2)\right)^{-1} = q^2 D(q^2)\left[
\frac{\log(\frac{q^2}{\Lambda^2})}{\log(\frac{\mu^2}{\Lambda^2})}\right]^{d_D},
\end{eqnarray}
where $d_D = \frac{13}{22}$ for the quenched approximation in
Landau gauge, $a\mu=1$ by choice, and $\Lambda = 0.35(5)$ is
obtained from a fit to the propagator. We note that the error is
obtained from the sensitivity of the result to the upper limit of
this fit range in the region $0.5\le (aq)^2 \le 2.0$. Our value
for $\Lambda a$ is identical to that obtained in
Ref.~\cite{special_gluon} when fitted over the range used in that
paper, but it shows great sensitivity to the range.

\begin{figure}[hbt]
\begin{center}
\includegraphics[width=1.0\columnwidth,angle=0]{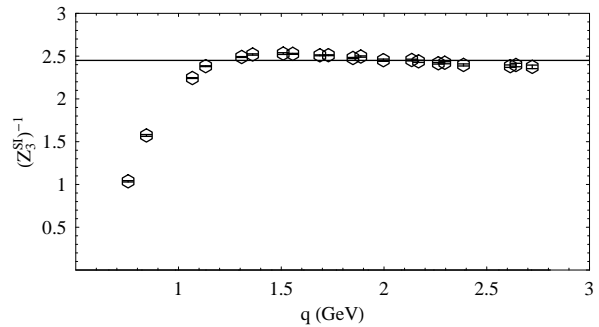}
\end{center}
\caption{ \label{fit_gprop} Scale invariant MOM scheme
renormalization constant $Z_3^\mathrm{SI}(q^2)^{-1}$ at
\emph{fixed} renormalization point $(a\mu)^2=1$. }
\end{figure}

We can see that this scale invariant presentation in
Figure~\ref{fit_gprop} 
shows a nice plateau demonstrating perturbative behavior and gives
confidence in our result at and around this renormalization point.
We obtain the MOM scheme renormalization constant as
$(Z_3^\mathrm{SI})^{-1} = 2.45(1)$ at the scale of $\mu = a^{-1} =
1.922(40)\mathrm{\ GeV}$.

A more conventional approach would be to use the
$\overline{\mathrm{MS}}$ scheme at some scale.  However, the
conversion to this scheme is not known and we isolate this lack of
knowledge, localizing it to a single continuum renormalization
constant $\overline{Z_3}(\overline{\mu})$ which converts from the
RI-MOM scheme at $a\mu = 1$ to the $\overline{\mathrm{MS}}$ scheme
at momentum scale $\overline{\mu}$.

\section {Quark-quark-gluon three-point function}
\begin{figure}[hbt]
\includegraphics[width=\columnwidth]{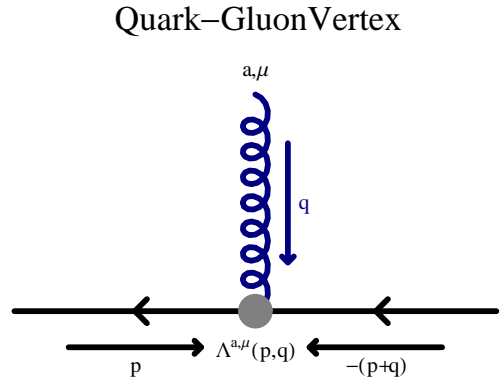}
\caption{\label{fig3point} Quark-quark-gluon momentum space,
three-point function.}
\end{figure}

We consider the three-point function consisting of a gluon with
incoming momentum $q$ and two quarks with incoming momentum $p$
and $-(p+q)$ as shown in Figure~\ref{fig3point}.

\label{SecVertex}
\subsection{Lorentz structure of vertex}

In the continuum the vertex function has been expressed in terms
of its general Lorentz structure in Ref.~\cite{qgv}. This
decomposition is complex and requires a comprehensive set of
lattice momenta to resolve the different components.

In the following subsections we will study kinematic regions that
simplify the form factors appearing in the decomposition allowed
by Lorentz symmetry in each case. As usual, appropriate spin and
momentum projections can be used to determine the form factors
from the bare lattice data.

\subsection{Hard recoil kinematics}

We first require both quark lines to carry momentum $p$ giving a gluon
momentum of $-2p$.  In this case the vertex function may be decomposed as follows
\begin{eqnarray}
\Lambda_{\mu}^{P}(p,q=-2p) &=&
-ig\left[\lambda_1^{\prime}(q^2)\left(\gamma_{\mu}-{\ooalign{\hfil/\hfil\crcr$q$}}q_{\mu}/q^2\right)
\right.\nonumber \\
&-& \left. i \tau_5(q^2)\sigma_{\mu\nu}q_{\nu}
\vphantom{\lambda_1^{\prime}}\right].
\end{eqnarray}
The form factors $\lambda_1^{\prime}$ and $\tau_5$ can be easily
separated by different Lorentz projections.

\subsection{Soft gluon kinematics}

Next we choose the point $q=0$.  At this kinematic point the
tensor contribution vanishes, and we may write these form factors
in terms of their popular Lorentz decomposition as follows:
\begin{eqnarray}
\Lambda_{\mu}^{}(p,q=0) &=& -ig\left[\lambda_1(p^2)\gamma_{\mu} -
4\lambda_2(p^2) {\ooalign{\hfil/\hfil\crcr$p$}}p_{\mu}\right.\nonumber\\
&-&\left. 2i\lambda_3(p^2)p_{\mu}\right].
\end{eqnarray}
The form factors $\lambda_1$ and $\lambda_2$ are coupled by the
same Lorentz projections but can be decoupled as following:
\begin{eqnarray}
\lambda_1(p^2) &=& \frac{1}{36} \sum_\mu
\mathrm{Tr}\left(\gamma_\mu \sum_\nu\left(\delta_{\mu,\nu} -
\frac{p_\mu
p_\nu}{p^2}\right)\Lambda_{\nu}^{\mathrm{lat}}\right)\\
\lambda_2(p^2) &=& \frac{1}{48(pa)^2} \left[\sum_\mu
\mathrm{Tr}\left(\gamma_\mu
\Lambda_{\mu}^\mathrm{lat}\right) -\frac{4}{3}\lambda_1(p^2)\right]\\
\lambda_3(p^2) &=& \frac{i}{24(pa)^2} \sum_\mu
\mathrm{Tr}\left(p_\mu \Lambda_{\mu}^\mathrm{lat}\right).
\end{eqnarray}
Our method projects $\lambda_1$ and $\lambda_2$ for general
kinematics while in Ref.~\cite{qgv} the authors use only kinematic
points with $p_\mu = 0$ to isolate $\lambda_1$ from $\lambda_2$.

\subsection{Tree-level lattice perturbation theory} %
Here we note that the tree-level vertex function for the clover
lattice action (without off-shell improvements) is
\begin{eqnarray}
\Lambda_{\mu}^{[0], \mathrm{lat}} &=&\gamma_{\mu} \cos\left(p_{\mu}a+\frac{1}{2}q_{\mu}a\right)
 - i \sin\left(p_{\mu}a+\frac{1}{2}q_{\mu}a\right)\nonumber\\
  & + & \frac{1}{2}c_\mathrm{SW}\cos\left(\frac{1}{2}q_{\mu}a\right)
  \sum_{\mu \neq \nu}\sigma_{\mu\nu}\sin \left(q_{\nu}a\right).
\end{eqnarray}
Thus, we might expect that to leading-order the clover action will
be most sensitive to the clover coefficient through terms
proportional to $\sigma_{\mu\nu}$.  Therefore, in investigating
the feasibility of fixing the clover coefficient from the vertex
function we will consider the vertex function defined from bare
(un-rotated) quark fields. We shall see later that, somewhat
interestingly, this turned up a negative result.

Note that in subsection~\ref{form_factors}, we will use the
``bare'' clover action to calculate the various form factors to
avoid contamination from the poor choice of off-shell improved
coefficients, while in subsection~\ref{cq_prime}, we will focus on
the off-shell improved vertex with the non-perturbatively improved
clover action.

\subsection{Form factors}
\label{form_factors} 
\begin{figure}[hbt]
\includegraphics[width=1.0\columnwidth]{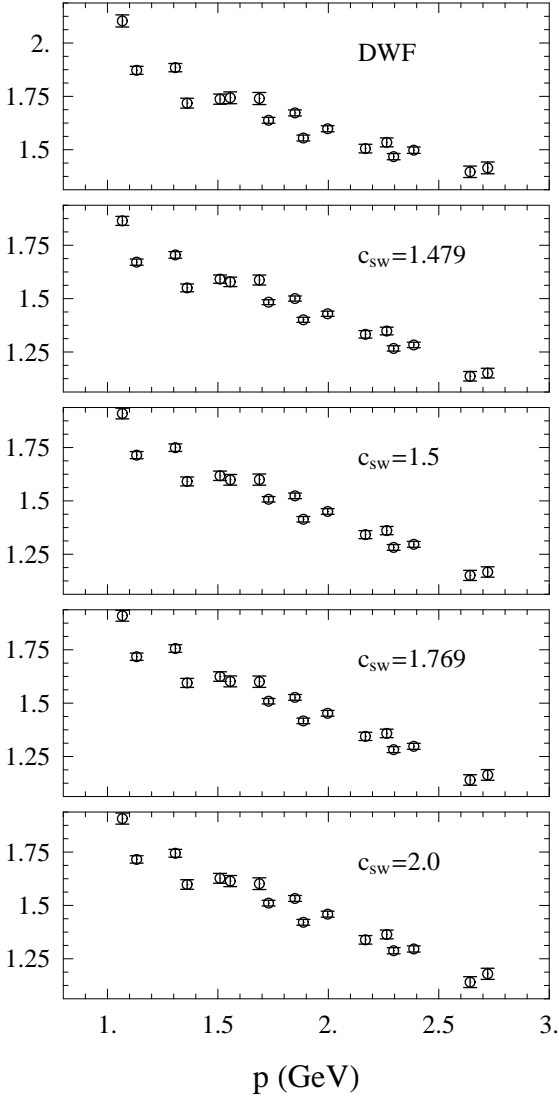}
\vspace{-3.0cm}%
 \caption{ \label{fig_lambda_1} The vector vertex function form
factor $\lambda_1(p^2)/\overline{Z_3}(\overline{\mu})$ for the
soft gluon case. This is displayed for both the chirally symmetric
DWF action and the bare clover action.}
\end{figure}

\begin{figure}[hbt]
\includegraphics[width=1.0\columnwidth]{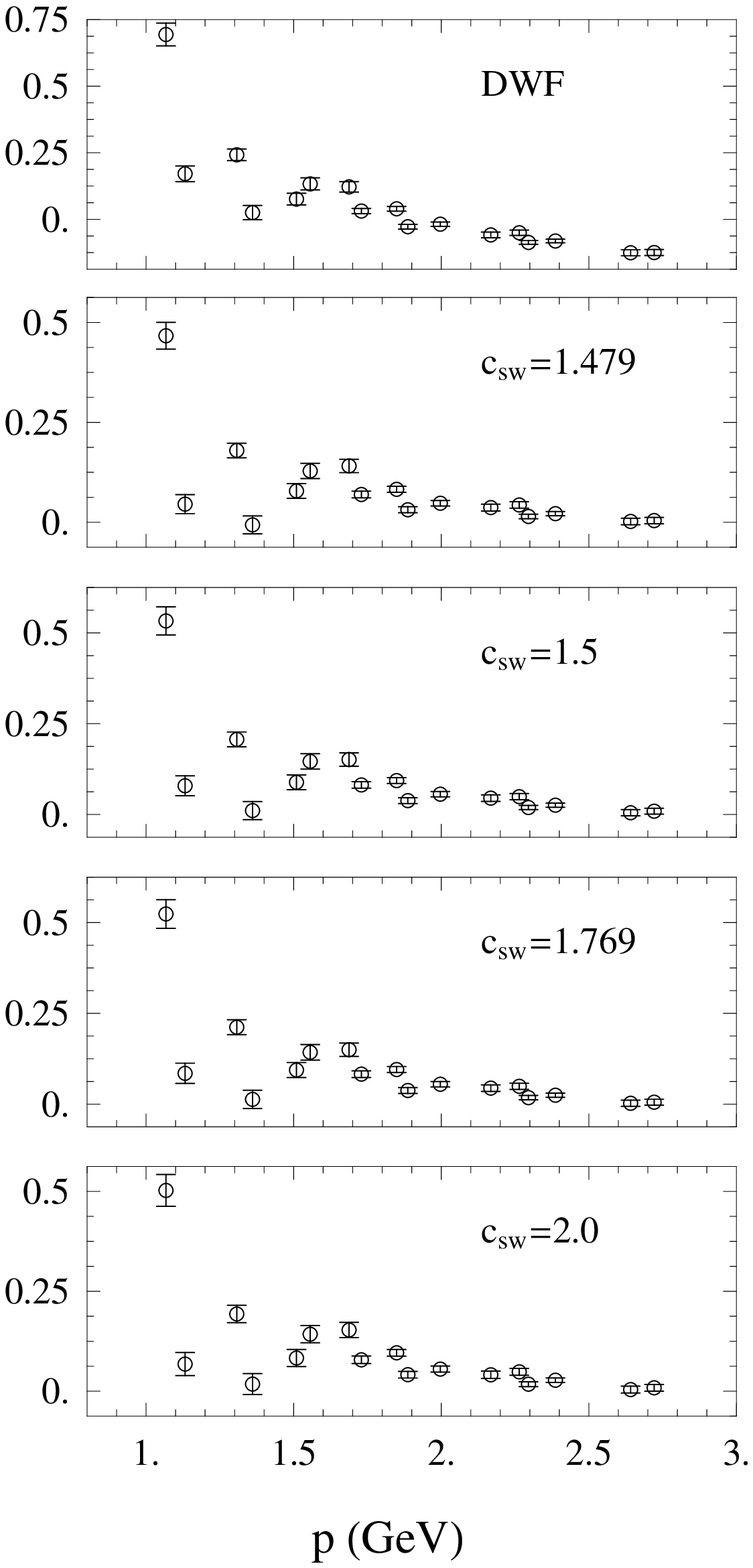}
\vspace{-3.0cm}%
\caption{ \label{fig_lambda_2} The vector vertex function form
factor $\lambda_2(p^2)/\overline{Z_3}(\overline{\mu})$ for the
soft gluon case. This is displayed for both the chirally symmetric
DWF action and the bare clover action.}
\end{figure}

Figures~\ref{fig_lambda_1}-\ref{fig_tau_5}
display the lattice results for $\lambda_1(p^2)$,
$\lambda_2(p^2)$, $\lambda_3(p^2)$, $\lambda_1^\prime(p^2)$, and
$\tau_5(p^2)$ in the $\overline{\mathrm{MS}}$ scheme using the
renormalization constants calculated earlier in this paper, up to
the unknown constant $\overline{Z_3}(\overline{\mu})$ in
Section~\ref{SecGluonProp}. This is shown for both the bare clover
actions and the domain wall action. We will discuss the off-shell
improvement of the clover action via post-simulation rotations in
the following section.

\begin{figure}[hbt]
\includegraphics[width=1.0\columnwidth]{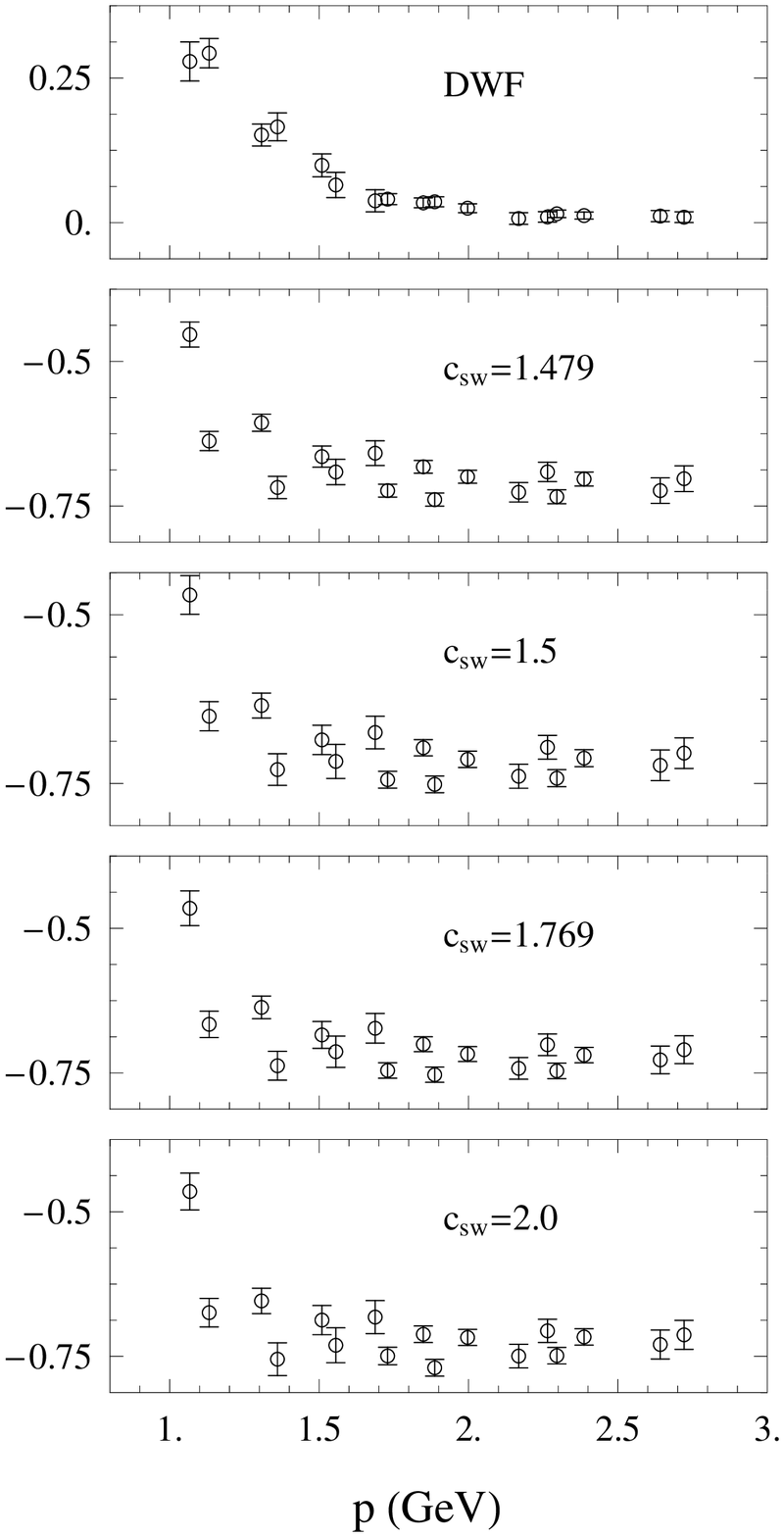}
\vspace{-3.0cm}%
\caption{ \label{fig_lambda_3} The scalar vertex function form
factor $\lambda_3(p^2)/\overline{Z_3}(\overline{\mu})$ for the
soft gluon case. This is displayed for both the chirally symmetric
DWF action and the bare clover action.}
\end{figure}

We note that the off-shell improved DWF action in general shows a
marked difference in the $\lambda_3$ scalar form-factor; see
Figure~\ref{fig_lambda_3}, in particular. The difference between the DWF
and clover action results is likely to be caused by the off-shell improvement
present with domain wall fermions,
and one might hope that the form factor $\lambda_3$  would be
a good means to adjust the off-shell improvement coefficients for
clover actions. There will be more discussion in Section~\ref{cq_prime}
on this possibility.

\begin{figure}[hbt]
\includegraphics[width=1.0\columnwidth]{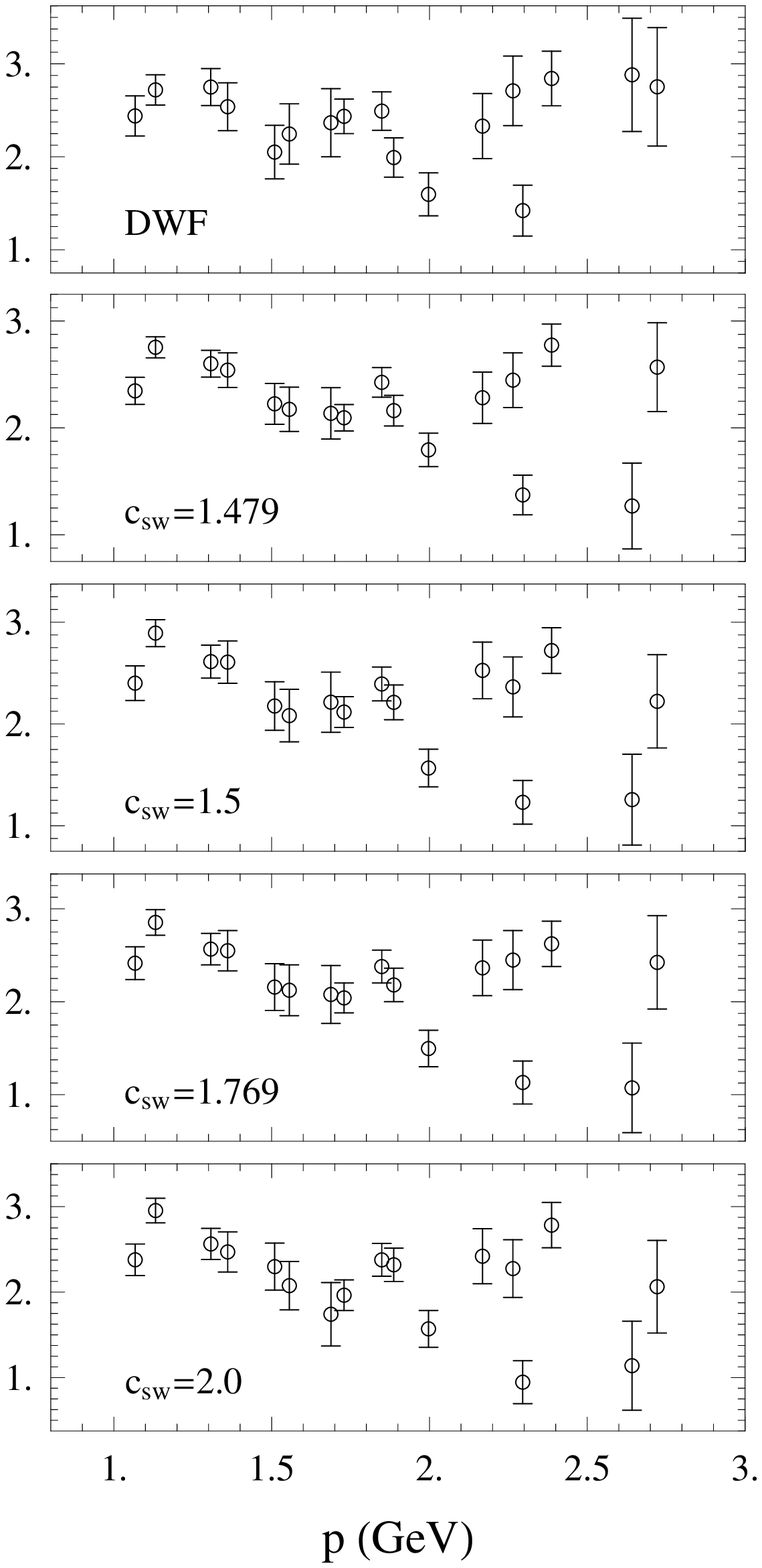}
\vspace{-3.0cm}%
\caption{ \label{fig_lambda_3} The vector vertex function form
factor $\lambda_1^{\prime}(p^2)/\overline{Z_3}(\overline{\mu})$
for the hard-recoil kinematics. This is displayed for both the
chirally symmetric DWF action and the bare clover action. }
\end{figure}

The quantity $\lambda_1^{\prime} $ is the form factor reflecting the
$\gamma^\mu$ term in the fermion action. This quantity appears to be
the most similar among the actions studied here.

\begin{figure}[hbt]
\includegraphics[width=1.0\columnwidth]{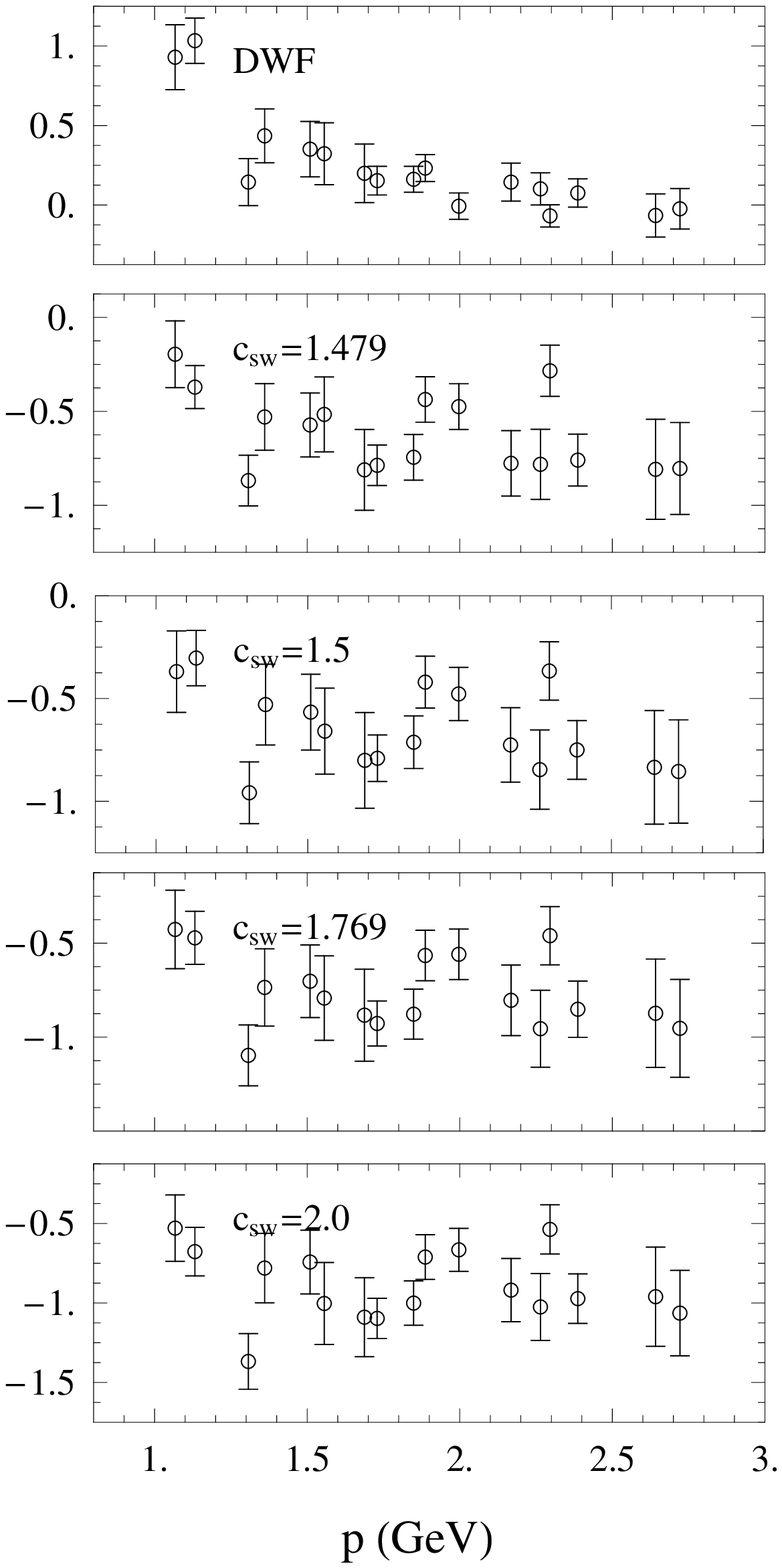}
\vspace{-3.0cm}%
\caption{ \label{fig_tau_5} The tensor vertex function form factor
$\tau_5(p^2)/\overline{Z_3}(\overline{\mu})$ for the hard-recoil
kinematics. This is displayed for both the chirally symmetric DWF
action and the bare clover action.  }
\end{figure}

As seen in Figure~\ref{fig_tau_5}, we find no statistically
significant evidence of $c_\mathrm{SW}$ dependence in the $\sigma$
projection of the vertex function, despite both its leading order
contribution to the tree-level vertex and our examining it at a
sizeable range of lattice momenta.

The difference between the (bare) clover and DWF actions is again
quite marked and may be in principle useful for determining
off-shell improvement rotations. However, since $c_\mathrm{SW}$
(in perturbation theory) couples with this vertex function at
leading order and since the non-zero gluon momenta are
statistically noisier, we will only calculate the $c_q^\prime$
from $\lambda_3$.

\subsection{Determining $c_q^\prime$ directly from the vertex function}
\label{cq_prime} The quantity $c_q^\prime$ appearing in the
improved quark field can be calculated from the quark propagator
alone, as demonstrated previously in
Refs.~\cite{Martinelli_offShell,C_ngi} and in this work. Here we
propose and will demonstrate that we can determine this
coefficient independently by matching form factors to the already
improved DWF vertex function for the $q = 0$ kinematics. Any
consistency between these two independent determinations of
$c_q^\prime$ will contribute to the body of evidence that
$c_\mathrm{NGI}$ is near zero.

\subsubsection{Tree-level calculation}%
If we apply the off-shell quark field improvement of
Eq.(\ref{eq:off_shell_improv_psi}), we obtain the following
expression for the tree-level vertex function\cite{hwlin04}:
\begin{eqnarray}
\Lambda_{\mu}^{[0]}
& = &%
 \left\{i\sin(p_{\mu}a+q_{\mu}a/2)(-1+ 2c_q) \right. \nonumber\\
\hskip -0.2in &+&
\left.ic_\mathrm{NGI}\left[\sin\left(p_{\mu}a+q_{\mu}a\right)
 + \sin\left(p_{\mu}a\right)\right]\cos \left(p_{\mu}a+q_{\mu}a/2\right)\right\} \nonumber\\
&-& \gamma_{\mu} \left\{\cos \left(p_{\mu}a+q_{\mu}a/2\right)(1 + 4c_q ma)\right\}\nonumber\\
&+& i\sigma_{{\mu}{\nu}}\left\{{i}/{2}\cos \left(q_{\mu}a/2\right)
\sin\left(q_{\nu}a\right)
\left(c_\mathrm{SW} -2c_q\right) \right. \nonumber\\
&+ & \left.
ic_\mathrm{NGI}\left[\sin\left(p_{\nu}a+q_{\nu}a\right)-
\sin\left(p_{\nu}a\right)\right]\cos \left(p_{\mu}a+q_{\mu}a/2\right) \right\}\nonumber\\
\end{eqnarray}
which depends on the coefficients of $c_q^\prime$ (=
$-$0.5$c_q$) and $c_\mathrm{NGI}$.  As above we assume
$c_\mathrm{NGI}$ is zero, so the only task is to determine the
quantity $c_q^\prime$ from the vertex function.

This expression illuminates how the different form factors might
be expected to depend (to leading order) on the different
improvement coefficients. In particular, we note the leading-order
sensitivity of the $\sigma_{\mu\nu}$ piece, $\tau_5$,
to the clover coefficient. %

\subsubsection{Numerical work}%
Here we consider the clover action with the non-perturbative
$c_\mathrm{SW}$ coefficient. We included the $ma$ dependence (from
on-shell improvement) by replacing the $Z_q$ in
Eq.(\ref{eq:off_shell_improv_psi}) with $(1 + b_q ma)Z_q $ for the
clover action and performed the same calculation steps as in the
unimproved field cases. Since the $ma$ is small and $b_q$ is 1.0
at tree level, we expected the contribution from $b_q$ to be
negligible. Indeed, we saw no significant changes in the
calculation on quark propagator and form factors from quark-gluon
vertex function.

\begin{figure}[!hbt]
\vspace{0cm}
 \begin{tabular}{c}
\includegraphics[width=0.95\columnwidth]{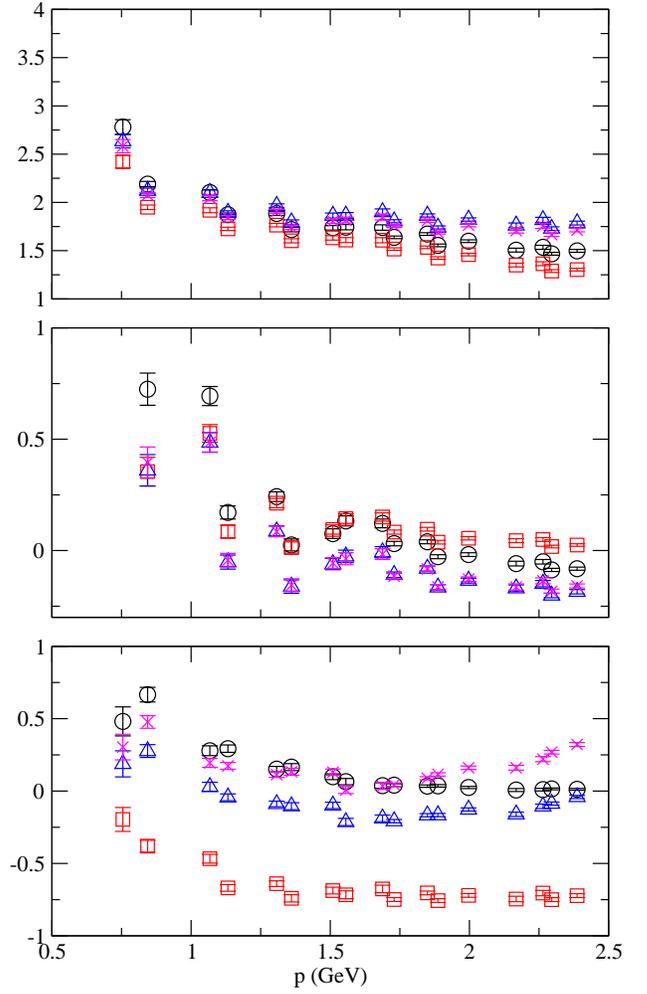}\\
\end{tabular}
\caption{ \label{bq1.0_gamma} The form factors $\lambda_1$,
$\lambda_2$, $\lambda_3$ (from top to bottom) calculated with $q =
0$ kinematics and with various $c_q^\prime$. The various actions
are represented by different symbols: DWF (circles); NP clover
with $c_q^\prime= 0.0$ (squares), $c_q^\prime = -0.20$ (triangles)
and $c_q^\prime= -0.257$ (crosses). }
\end{figure}

In Figure~\ref{bq1.0_gamma} we compare the three form factors
$\lambda_1$, $\lambda_2$ and $\lambda_3$ between the domain wall
and bare clover actions where the latter are computed for various
values of $c_q^\prime \in \{0.0, -0.20, -0.257\}$ and with $b_q$
taking its tree-level value, $b_q$ = 1.0. As we
predict from our tree-level exercises, the form factor $\lambda_3$
is indeed very sensitive to $c_q^\prime$.  The calculation shows
that our value for $c_q^\prime$ calculated from quark propagator
is closest to the DWF result, although  at large momentum, the
$(pa)^2$ effects dominate and disagree with the completely
off-shell improved result. Clearly we successfully confirm an
independent method to determine $c_q^\prime$.

\section{Discussion }
\label{SecDiscuss} We have calculated the form factors
$\lambda_1$, $\lambda_2$, $\lambda_3$, $\lambda_1^\prime$ and
$\tau_5$ for both the bare clover action and the DWF action in two
kinematic regions. We have numerically demonstrated rather clearly
the automatic off-shell improvement of chiral fermion
formulations.

Performing post-simulation rotations on the clover data can yield
improved form factors. This involves worrisome systematic errors,
such as the taking of $c_\mathrm{NGI} = 0$, which are entirely
avoided in the DWF formulation.

We demonstrate here that the use of the vertex function form
factors as a means of fixing at least one improvement coefficient
of the clover action in a manner that does not require the light
quark limit. Unfortunately we have not been able to fix the
crucial on-shell coefficient $c_\mathrm{SW}$.

From the vertex form factors for the $q = 0$ kinematics, we showed
that we can provide an independent way of determining the
coefficient $c_q^\prime$ appearing in the improved quark field for
the non-perturbatively improved clover action. Our finding of
$c_q^\prime=-0.257$ is very close to the tree-level
prediction\cite{Martinelli_offShell}, $-0.25$ with the same
tree-level $b_q$, or to the mean-field value of $-0.285$ with $b_q
= 1.14$, and also to previous calculations\cite{C_ngi} from the
quark propagator alone.

We note that simulations with different clover coefficients and
matched pseudoscalar masses produced identical vertex function
form factors. One would na\"ively expect to see a difference due
to different choices of clover coefficient, particularly in the
$\tau_5$ form factor and in the large momentum region where the
leading perturbative behavior should dominate and the universal
cut-off-insensitive infrared physics disappears. However, it seems
that the we have not yet reached a sufficiently fine lattice
spacing where this perturbative intuition applies. Thus this
coefficient cannot be fixed at these lattice spacings using this
approach.

In contrast to the insensitivity to the clover coefficient, we see
a large change in off-shell behavior between the entire clover
class of actions and the domain wall action. In particular, since
the domain wall action is \(O(a)\) off-shell improved and this
(un-rotated) clover action is not, large differences serve to
demonstrate the effectiveness of the domain wall action, and it
surely is a better action to study the vertex function in the
future.

\section{Summary and Outlook }
\label{SecFuture} We have studied the non-perturbative structure
of two- and three-point functions for both the domain wall fermion
action with an automatically \(O(a)\) off-shell improved quark
field, and for the on-shell improved clover action with various
choices of the $c_\mathrm{SW}$ coefficient at fixed pseudoscalar
mass. For the clover action the vertex function shows weak
dependence on various choices of $c_\mathrm{SW}$ at fixed
pseudoscalar mass. Studying various Dirac structures at large
momentum for the domain wall action showed vast improvements in
the off-shell behavior of the theory. Also promising is the use of
the behavior of the vertex function as a new method of fixing the
\(O(a)\) off-shell improved field coefficients for other fermion
actions, in particular during the non-perturbative step scaling of
relativistic heavy quark actions.

Comparison with the less-improved clover action allowed us to
provide a new method to determine the coefficient $c_q^\prime$
through the form factor $\lambda_3$.   The result found was
consistent with the value calculated from the quark propagator and
previous published results. Further, this consistency may be taken
as additional evidence that $c_\mathrm{NGI}$ is negligible.

\section*{ACKNOWLEDGMENTS}
\vspace{-0.05in} The author would like to thank RIKEN, Brookhaven
National Laboratory and the U.S. Department of Energy for
providing the facilities essential for the completion of this work
and N. Christ, P. Boyle, C. Dawson, T. Izubuchi for useful
discussions on various subjects and J. Skullerud for kindly
providing
both a vertex function calculation comparison and useful conversations.%


\vspace{-0.1in}



\begin{thebibliography}{1}
\expandafter\ifx\csname bibnamefont\endcsname\relax
  \def\bibnamefont#1{#1}\fi
\expandafter\ifx\csname bibfnamefont\endcsname\relax
  \def\bibfnamefont#1{#1}\fi
\expandafter\ifx\csname url\endcsname\relax
  \def\url#1{\texttt{#1}}\fi
\expandafter\ifx\csname
urlprefix\endcsname\relax\def\urlprefix{URL }\fi
\expandafter\ifx\csname bibinfo\endcsname\relax
\def\bibinfo#1#2{#2}\fi
\expandafter\ifx\csname eprint\endcsname\relax
\def\eprint#1{#1}\fi


\bibitem{NP_CLOVER}
\bibinfo{author}{\bibfnamefont{M.~Luscher}}, \emph{et~al.},
  \bibinfo{journal}{Nucl. Phys. B} \textbf{\bibinfo{volume}{491}},
  \bibinfo{pages}{323} (\bibinfo{year}{1997}),
  \eprint{hep-lat/9609035};
\bibinfo{author}{\bibfnamefont{M.~Luscher}}, \emph{et~al.},
  \bibinfo{journal}{Nucl. Phys. B} \textbf{\bibinfo{volume}{491}},
  \bibinfo{pages}{344} (\bibinfo{year}{1997}),
  \eprint{hep-lat/9611015}.

\bibitem{Fermilab_action}
\bibinfo{author}{\bibfnamefont{A.~X.~El-Khadra}}, \emph{et~al.},
  \bibinfo{journal}{Phys. Rev. D} \textbf{\bibinfo{volume}{55}},
  \bibinfo{pages}{3933} (\bibinfo{year}{1999}),
  \eprint{hep-lat/9604004}.


\bibitem{quenched_g_prop}
\bibinfo{author}{\bibfnamefont{J.~E.~Mandula}}, \emph{et~al.},
  \bibinfo{journal}{Phys. Lett. B} \textbf{\bibinfo{volume}{185}},
  \bibinfo{pages}{127} (\bibinfo{year}{1987});
\bibinfo{author}{\bibfnamefont{C.~W.~Bernard}}, \emph{et~al.},
  \bibinfo{journal}{Phys. Rev. D} \textbf{\bibinfo{volume}{49}},
  \bibinfo{pages}{1585} (\bibinfo{year}{1994}),
  \eprint{hep-lat/9307001};
\bibinfo{author}{\bibfnamefont{P.~Marenzoni}}, \emph{et~al.},
  \bibinfo{journal}{Nucl. Phys. B} \textbf{\bibinfo{volume}{455}},
  \bibinfo{pages}{339} (\bibinfo{year}{1995}),
  \eprint{hep-lat/9410011};
\bibinfo{author}{\bibfnamefont{D.~B.~Leinweber}}, \emph{et~al.},
  \bibinfo{journal}{Phys. Rev. D} \textbf{\bibinfo{volume}{58}},
  \bibinfo{pages}{031501} (\bibinfo{year}{1998}),
  \eprint{hep-lat/9803015};
\bibinfo{author}{\bibfnamefont{D.~Becirevic}}, \emph{et~al.},
  \bibinfo{journal}{Phys. Rev. D} \textbf{\bibinfo{volume}{60}},
  \bibinfo{pages}{094509} (\bibinfo{year}{1999}),
  \eprint{hep-ph/9903364};
\bibinfo{author}{\bibfnamefont{J.~P.~Ma}},
  \bibinfo{journal}{Mod. Phys. Lett. A} \textbf{\bibinfo{volume}{15}},
  \bibinfo{pages}{229} (\bibinfo{year}{2000}),
  \eprint{hep-lat/9903009};
\bibinfo{author}{\bibfnamefont{D.~Becirevic}}, \emph{et~al.},
  \bibinfo{journal}{Phys. Rev. D} \textbf{\bibinfo{volume}{61}},
  \bibinfo{pages}{114508} (\bibinfo{year}{2000}),
  \eprint{hep-ph/9910204};
\bibinfo{author}{\bibfnamefont{H.~Nakajima}}, \emph{et~al.},
  \bibinfo{journal}{Nucl. Phys. A} \textbf{\bibinfo{volume}{680}},
  \bibinfo{pages}{151} (\bibinfo{year}{2000}),
  \eprint{hep-lat/0004023};
\bibinfo{author}{\bibfnamefont{F.~D.~R.~Bonnet}}, \emph{et~al.},
  \bibinfo{journal}{Phys. Rev. D} \textbf{\bibinfo{volume}{62}},
  \bibinfo{pages}{051501} (\bibinfo{year}{2000}),
  \eprint{hep-lat/0002020};
\bibinfo{author}{\bibfnamefont{F.~D.~R.~Bonnet}}, \emph{et~al.},
  \bibinfo{journal}{Phys. Rev. D} \textbf{\bibinfo{volume}{64}},
  \bibinfo{pages}{034501} (\bibinfo{year}{2001}),
  \eprint{hep-lat/0101013};
\bibinfo{author}{\bibfnamefont{K.~Langfeld}}, \emph{et~al.},
  \bibinfo{journal}{Nucl. Phys. B} \textbf{\bibinfo{volume}{621}},
  \bibinfo{pages}{131} (\bibinfo{year}{2002}),
  \eprint{hep-ph/0107141};
\bibinfo{author}{\bibfnamefont{P.~O.~Bowman}}, \emph{et~al.},
  \bibinfo{journal}{Phys. Rev. D} \textbf{\bibinfo{volume}{66}},
  \bibinfo{pages}{074505} (\bibinfo{year}{2002}),
  \eprint{hep-lat/020601}.

\bibitem{quark_prop}
\bibinfo{author}{
\bibnamefont{C.~W.~Bernard}, \emph{et~al.}},
  \bibinfo{journal}{Nucl. Phys. B (Proc. Suppl.)} \textbf{\bibinfo{volume}{17}},
  \bibinfo{pages}{593} (\bibinfo{year}{1990});
\bibinfo{author}{\bibfnamefont{C.~W.~Bernard}}, \emph{et~al.},
  \bibinfo{journal}{Nucl. Phys. B (Proc. Suppl.)} \textbf{\bibinfo{volume}{20}},
  \bibinfo{pages}{410} (\bibinfo{year}{1991});
\bibinfo{author}{\bibfnamefont{J.~I.~Skullerud}}, \emph{et~al.},
  \bibinfo{journal}{Phys. Rev. D} \textbf{\bibinfo{volume}{63}},
  \bibinfo{pages}{054508} (\bibinfo{year}{2001});
\bibinfo{author}{\bibfnamefont{J.~I.~Skullerud}}, \emph{et~al.},
  \bibinfo{journal}{Phys. Rev. D} \textbf{\bibinfo{volume}{64}},
  \bibinfo{pages}{074508} (\bibinfo{year}{2001});
\bibinfo{author}{\bibfnamefont{J.~I.~Skullerud}}, \emph{et~al.},
  \bibinfo{journal}{Nucl. Phys. B (Proc Suppl)} \textbf{\bibinfo{volume}{141}},
  \bibinfo{pages}{241} (\bibinfo{year}{2005});
\bibinfo{author}{\bibfnamefont{P.~O.~Bowman}}, \emph{et~al.},
  \bibinfo{journal}{Phys. Rev. D} \textbf{\bibinfo{volume}{66}},
  \bibinfo{pages}{014505} (\bibinfo{year}{2002});
\bibinfo{author}{\bibfnamefont{T.~Blum}}, \emph{et~al.},
  \bibinfo{journal}{Phys. Rev. D} \textbf{\bibinfo{volume}{66}},
  \bibinfo{pages}{014504} (\bibinfo{year}{2002});
\bibinfo{author}{\bibfnamefont{F.~D.~R.~Bonnet}}, \emph{et~al.},
  \bibinfo{journal}{Phys. Rev. D} \textbf{\bibinfo{volume}{65}},
  \bibinfo{pages}{114503} (\bibinfo{year}{2002});
\bibinfo{author}{\bibfnamefont{J.~B.~Zhang}}, \emph{et~al.},
  \bibinfo{journal}{Phys. Rev. D} \textbf{\bibinfo{volume}{70}},
  \bibinfo{pages}{034505} (\bibinfo{year}{2004});
\bibinfo{author}{\bibfnamefont{P.~O.~Bowman}}, \emph{et~al.},
  \bibinfo{journal}{Phys. Rev. D} \textbf{\bibinfo{volume}{70}},
  \bibinfo{pages}{034509} (\bibinfo{year}{2004}),
  \eprint{hep-lat/0402032}.

\bibitem{qgv}
\bibinfo{author}{\bibnamefont{J.~Skullerud}}, \emph{et~al.},
  \bibinfo{journal}{JHEP} \textbf{\bibinfo{volume}{0209}},
  \bibinfo{pages}{013} (\bibinfo{year}{2002}),
  \eprint{hep-ph/0205318};
\bibinfo{author}{\bibnamefont{J.~Skullerud}}, \emph{et~al.},
  \bibinfo{journal}{JHEP} \textbf{\bibinfo{volume}{0304}},
  \bibinfo{pages}{047} (\bibinfo{year}{2003}),
  \eprint{hep-ph/0303176};
\bibinfo{author}{\bibnamefont{P.~Boucaud}}, \emph{et~al.},
  \bibinfo{journal}{Phys.Lett B} \textbf{\bibinfo{volume}{575}},
  \bibinfo{pages}{256} (\bibinfo{year}{2003}),
  \eprint{hep-lat/0307026};
\bibinfo{author}{\bibnamefont{J.~Skullerud}}, \emph{et~al.},
  \bibinfo{journal}{Nucl. Phys. B (Proc. Suppl.)} \textbf{\bibinfo{volume}{128}},
  \bibinfo{pages}{117} (\bibinfo{year}{2004}).

\bibitem{DWF_prehistory}
\bibinfo{author}{\bibfnamefont{D.~B.~Kaplan}},
  \bibinfo{journal}{Phys. Lett. B} \textbf{\bibinfo{volume}{288}},
  \bibinfo{pages}{342} (\bibinfo{year}{1992}),
  \eprint{hep-lat/9206013};
\bibinfo{author}{\bibfnamefont{D.~B.~Kaplan}},
  \bibinfo{journal}{Phys. Proc. Suppl.} \textbf{\bibinfo{volume}{30}},
  \bibinfo{pages}{597} (\bibinfo{year}{1993});
\bibinfo{author}{\bibfnamefont{Y.~Shamir}},
  \bibinfo{journal}{Nucl. Phys. B} \textbf{\bibinfo{volume}{406}},
  \bibinfo{pages}{90} (\bibinfo{year}{1993}),
  \eprint{hep-lat/9303005};
\bibinfo{author}{\bibfnamefont{V.~Furman}}, \emph{et~al.},
  \bibinfo{journal}{Nucl. Phys. B} \textbf{\bibinfo{volume}{439}},
  \bibinfo{pages}{54} (\bibinfo{year}{1995}),
  \eprint{hep-lat/9405004}.

\bibitem{RBC_NPR}
\bibinfo{author}{\bibfnamefont{T.~Blum}}, \emph{et~al.} (\bibinfo{collaboration}{RBC}),
  \bibinfo{journal}{Phys. Rev. D} \textbf{\bibinfo{volume}{66}},
  \bibinfo{pages}{014504} (\bibinfo{year}{2002}),
  \eprint{hep-lat/0102005}.

\bibitem{hwlin04}
\bibinfo{author}{\bibnamefont{H.~Lin}} (\bibinfo{year}{2004}),
  \eprint{hep-lat/0409085}.

\bibitem{clover}
\bibinfo{author}{\bibfnamefont{Sheikholeslami and R.~Wohlert}},
  \bibinfo{journal}{Nucl. Phys. B} \textbf{\bibinfo{volume}{257}},
  \bibinfo{pages}{572} (\bibinfo{year}{1985}).

\bibitem{Martinelli_offShell}
\bibinfo{author}{\bibfnamefont{G.~Martinelli}}, \emph{et~al.},
  \bibinfo{journal}{Nucl. Phys. B} \textbf{\bibinfo{volume}{611}},
  \bibinfo{pages}{311} (\bibinfo{year}{2001}),
  \eprint{hep-lat/0106003};
\bibinfo{author}{\bibfnamefont{D.~Becirevic}}, \emph{et~al.},
  \bibinfo{journal}{Phys. Rev. D} \textbf{\bibinfo{volume}{61}},
  \bibinfo{pages}{114507} (\bibinfo{year}{2000}),
  \eprint{hep-lat/9909082}.

\bibitem{C_ngi}
\bibinfo{author}{\bibfnamefont{S.~R.~Sharpe}},
  \bibinfo{journal}{Nucl. Phys. B (Proc. Suppl.)} \textbf{\bibinfo{volume}{106}},
  \bibinfo{pages}{817} (\bibinfo{year}{2002}),
  \eprint{hep-lat/0110021};
\bibinfo{author}{\bibfnamefont{T.~Bhattacharya}}, \emph{et~al.}
  \bibinfo{journal}{Nucl. Phys. B (Proc. Suppl.)}  \textbf{\bibinfo{volume}{106}},
  \bibinfo{pages}{786} (\bibinfo{year}{2002}),
  \eprint{hep-lat/0111002};

\bibitem{wilson_g}
\bibinfo{author}{\bibfnamefont{Sheikholeslami and R.~Wohlert}},
  \bibinfo{journal}{Nucl. Phys. B} \textbf{\bibinfo{volume}{257}},
  \bibinfo{pages}{572} (\bibinfo{year}{1985}).

\bibitem{dwf_00}
\bibinfo{author}{\bibfnamefont{T.~Blum}}, \emph{et~al.},
  \bibinfo{journal}{Phys. Rev. D} \textbf{\bibinfo{volume}{69}},
  \bibinfo{pages}{074502} (\bibinfo{year}{2004}),
  \eprint{hep-lat/0007038}.

\bibitem{gribov}
\bibinfo{author}{\bibfnamefont{V.~N.~Gribov}},
  \bibinfo{journal}{Nucl. Phys. B} \textbf{\bibinfo{volume}{139}},
  \bibinfo{pages}{1} (\bibinfo{year}{1978}).

\bibitem{Yuri}
\bibinfo{author}{\bibfnamefont{Y.~Zhestkov}},
  \bibinfo{journal}{Ph.D. thesis (Columbia University)},  (\bibinfo{year}{2001}).

\bibitem{RI_NPR}
\bibinfo{author}{\bibfnamefont{G.~Martinelli}}, \emph{et~al.},
  \bibinfo{journal}{Nucl. Phys. B} \textbf{\bibinfo{volume}{445}},
  \bibinfo{pages}{81} (\bibinfo{year}{1995}),
  \eprint{hep-lat/9411010};
\bibinfo{author}{\bibfnamefont{V.~Gimenez}}, \emph{et~al.},
  \bibinfo{journal}{Nucl. Phys. B} \textbf{\bibinfo{volume}{531}},
  \bibinfo{pages}{429} (\bibinfo{year}{1998}),
  \eprint{hep-lat/9806006};
\bibinfo{author}{\bibfnamefont{A.~Doninir}}, \emph{et~al.},
  \bibinfo{journal}{Eur. Phys. J. C} \textbf{\bibinfo{volume}{10}},
  \bibinfo{pages}{121} (\bibinfo{year}{1999}),
  \eprint{hep-lat/9902030}.



\bibitem{gluon_fit}
\bibinfo{author}{\bibfnamefont{S.~Mandelstam}},
  \bibinfo{journal}{Phys. Rev. D} \textbf{\bibinfo{volume}{20}},
  \bibinfo{pages}{3223}(\bibinfo{year}{1979});
\bibinfo{author}{\bibfnamefont{G.~dell'Antonio}}, \emph{et~al.},
  \bibinfo{journal}{Nucl. Phys. B} \textbf{\bibinfo{volume}{326}},
  \bibinfo{pages}{333}(\bibinfo{year}{1989}).

\bibitem{special_gluon}
\bibinfo{author}{\bibfnamefont{D.~B.~Leinweber}}, \emph{et~al.},
  \bibinfo{journal}{Phys. Rev. D} \textbf{\bibinfo{volume}{60}},
  \bibinfo{pages}{094507} (\bibinfo{year}{1999}),
  \eprint{hep-lat/9811027}.


\end{thebibliography}
\end{document}